\documentstyle[prl,aps]{revtex}
\begin{document}
\twocolumn[\hsize\textwidth\columnwidth\hsize
\csname@twocolumnfalse%
\endcsname

\def\4he{$^4$He}
\def\pc{\protect\cite}
\def\hpsi{\hat\psi}
\def\tpsi{\tilde\psi}
\def\br{{\bf r}}
\def\brt{\br,t}
\def\bbrt{(\brt)}
\def\cphio{\Phi_0}
\def\beq{\begin{equation}}
\def\eeq{\end{equation}}
\def\bea{\begin{eqnarray}}
\def\eea{\end{eqnarray}}
\def\bna{\bbox{\nabla}}
\def\bp{{\bf p}}
\def\bv{{\bf v}}
\def\tn{\tilde n}
\def\tp{\tilde p}

\draft

\title{
Two-fluid hydrodynamics for a trapped weakly-interacting Bose gas}

\author{E. Zaremba}
\address{Department of Physics, Queen's University\\Kingston,
Ontario K7L 3N6, Canada} 

\author{and}
\address{}

\author{A. Griffin and T. Nikuni}
\address{Department of Physics, University of Toronto\\Toronto,
Ontario M5S 1A7, Canada}

\date{\today}

\maketitle

\begin{abstract}
We derive the coupled equations of motion for
the condensate (superfluid) and non-condensate (normal fluid)
degrees of freedom in a trapped Bose gas at finite temperatures.
Our results are based on the Hartree-Fock-Popov
approximation for the time-dependent condensate wavefunction 
and an assumption of local equilibrium for the non-condensate atoms.
In the case of a
uniform weakly-interacting gas, our formalism gives a microscopic
derivation of the well-known two-fluid equations of Landau.
The collective modes in a parabolically trapped Bose gas include
the
analogue of the out-of-phase second sound mode in uniform systems.

\end{abstract}

]

The low frequency dynamics of superfluid $^4$He is commonly
described using the two-fluid phenomenology
first developed by Tisza\protect\cite{tisza38} and 
Landau\protect\cite{landau41}. This description, later
shown to be a consequence of a Bose broken symmetry
\protect\cite{noz90,hoh65}, accounts for 
the characteristic features associated with superfluidity in terms 
of the relative motion of  normal fluid and superfluid degrees of
freedom. In
particular, it predicts the existence of second sound as an 
out-of-phase oscillation of the two components.  In the present
letter, we give a microscopic derivation of the analogous two-fluid
equations for an {\it inhomogeneous} weakly-interacting gas of trapped
atoms.  In this
situation, the superfluid is identified
with the condensate atoms as described by a macroscopic
wavefunction, while the normal fluid density is associated with the
non-condensate thermal cloud.  In the uniform density limit, we
show that our equations are consistent with the standard two-fluid
equations\protect\cite{landau41,noz90}.  However, the
hydrodynamic behavior of Bose gases is quite different from that
of a Bose liquid such as superfluid $^4$He.

Our analysis is based on the equation of motion of the
macroscopic Bose wavefunction \protect\cite{noz90,hoh65},
$\Phi({\bf r},t)$, as determined 
within the time-dependent Hartree-Fock-Popov
(HFP) approximation, which is a generalization of
Refs.\protect\cite{hut97} and \cite{griffin96}. 
This condensate wavefunction is coupled
%to the non-condensate density $\tilde n ({\bf r},t)$ of thermally
to the thermally excited atoms making up the non-condensate which 
we assume is described by a semi-classical phase space distribution
function $f({\bf r}, {\bf p}, t)$. The further assumption that
collisions are sufficiently rapid to force local equilibrium
within the thermal cloud leads to a set of hydrodynamic
conservation laws for the non-condensate component. Our final
result is a  set of  equations for the condensate and non-condensate 
degrees of freedom which extends recent work on the hydrodynamics of
trapped Bose gases\protect\cite{gri97} to below $T_{BEC}$.

We first consider the dynamics of the condensate. As usual, the
Bose 
field operator is conveniently separated into condensate and
non-condensate parts:
 $\hat\psi({\bf r}) =\Phi({\bf r})  +\tilde{\psi}({\bf r})$.
For an arbitrary nonequilibrium state, the spatially and
time-varying macroscopic wavefunction $\Phi({\bf r},t)
\equiv\langle\hat\psi({\bf r})\rangle_t$ is described within the
time-dependent Hartree-Fock-Popov approximation by the equation of
motion
\begin{eqnarray}
i\hbar{\partial\Phi({\bf r},t)\over\partial t}
= \Big [&-& {\hbar^2\nabla^2\over 2m} + U_{ext}({\bf r})
+ 2g\tilde n({\bf r},t) +gn_c({\bf r},t) \Big ] \nonumber \\
&\times& \Phi({\bf r},t) \equiv \hat {\cal H} ({\bf r},t) 
\Phi({\bf r},t)\ .
\label{eq1}
\end{eqnarray}

\noindent Here the nonequilibrium non-condensate density is given
by $\tilde n({\bf r},t) =\langle\tilde\psi ^\dagger ({\bf r})
\tilde\psi({\bf r})\rangle_t$, the condensate density 
is $n_c({\bf r},t) =\vert\Phi({\bf r},t)\vert^2$ and $g=4\pi
a\hbar^2/m$ is the interaction strength. Eq.(\ref{eq1}) 
represents a natural
extension of  recent work  \protect\cite{hut97,griffin96} which
approximated $\tilde n ({\bf r},t)$ in (\ref{eq1}) by the equilibrium
value $\tilde n_0({\bf r})$, thereby ignoring the collective 
behavior of the non-condensate.  In contrast, the time-dependent 
condensate wavefunction $\Phi({\bf r},t)$ in (\ref{eq1})  is 
coupled into the fluctuations of the non-condensate and a dynamical
equation for the latter is also required.

It is convenient to recast the condensate equation of motion into
a pair
of hydrodynamic equations using the amplitude and phase
representation
$\Phi({\bf r},t) =\vert\Phi({\bf r},t)\vert e^{i\theta({\bf
r},t)}$.
Substituting this form into (\ref{eq1}) and separating real and
imaginary parts, one finds
\begin{eqnarray}
{\partial n_c\over \partial t} &=& -\bbox{\nabla} \cdot
(n_c{\bf v}_c) \nonumber \\
m\Bigg [{\partial {\bf v}_c\over \partial t} &+&{1\over
2}\bbox{\nabla}{\bf v}_c^2 \Bigg ] = -\bbox{\nabla}
\phi\ ,
\label{eq3}
\end{eqnarray}

\noindent
where ${\bf v}_c ({\bf r},t) \equiv \hbar\bbox{\nabla}\theta({\bf
r},t)/m$.  The potential $\phi({\bf r},t)$ is defined by

\begin{equation}
\phi({\bf r},t)\equiv {1\over \vert\Phi({\bf r},t)\vert} \hat{\cal
H}({\bf r},t) \vert
\Phi({\bf r},t)\vert\ ,
\label{eq4}
\end{equation}
where $\hat {\cal H}({\bf r},t)$ is the HFP Hamiltonian 
given in (\ref{eq1}). Anticipating the identification of ${\bf
v}_c$
with the superfluid velocity ${\bf v}_S$, we see that $\phi$ plays
the role of the chemical potential associated with the superfluid
motion\cite{noz90}. 
%The equations in (\ref{eq3}) emphasize how the fluctuations
%of the amplitude and phase of $\Phi({\bf r},t)$ are coupled.

We next consider the dynamics of the non-condensate in the
low-frequency collision-dominated hydrodynamic regime. 
In the semi-classical limit valid at finite temperatures\cite{gio} 
(with $k_BT\gg \hbar\omega_0,\, gn_0({\bf r})$, where
$\omega_0$ is a characteristic trap frequency), the
dynamics can be formulated in terms of a quantum kinetic equation for
the distribution function $f({\bf r},{\bf p},t)$ 
\protect\cite{KB,ref11}.    
Since (\ref{eq3}) implies that the number of particles in
the condensate is conserved, we must for consistency exclude those
processes which scatter atoms in and out of the condensate\cite{ref11}. 
In this
situation, only collisions between excited atoms are relevant and we can
use the kinetic equation\protect\cite{KB}
%leads to a separate equation of continuity for
%the condensate as given in (\ref{eq3}), the non-condensate should 
%also satisfy such a continuity equation.  We conclude that for 
%consistency, the use of the HFP approximation in (\ref{eq1}) means 
%that our kinetic equation for the non-condensate can only include 
%collisions between excited atoms (i.e. no collisions involving 
%taking atoms in/out of condensate).  In turn, this means the 
%kinetic equation for $f({\bf p}, {\bf r},t$) can be formulated in 
%terms of atoms \protect\cite{KB}, rather than in terms of 
%excitations as is done in  Ref.\protect\cite{ref11}.  Our 
%analysis is based on
\begin{equation}
\left [{\partial\over \partial t} + {{\bf p} \over m}\cdot
\bbox{\nabla}_r
- \bbox{\nabla} U({\bf r},t)\cdot\bbox{\nabla}_p\right ] 
f({\bf r},{\bf p},t) = \left. {\partial
f\over \partial t}\right\vert_{coll}
\label{eq5}
\end{equation}

\noindent Here $U({\bf r},t) \equiv U_{ext}({\bf r}) + 2g [\tilde
n({\bf r},t)+
n_c({\bf r},t)]$ includes the self-consistent Hartree-Fock dynamic
mean field in which the condensate part $2gn_c({\bf r},t)$ can be
viewed
as an additional external field acting on the
non-condensate.

The required hydrodynamic equations are obtained from 
(\ref{eq5}) by making
the further assumption that collisions enforce the distribution
function
to take the local equilibrium form\protect\cite{KB}

\begin{equation}
f_0({\bf r},{\bf p},t) = {1\over {\mathrm exp} \left [ \beta\{ 
{1\over 2m} [{\bf p}-m{\bf v}_n]^2
+ U-\mu\}\right ] -1} \ ,
\label{eq6}
\end{equation}

\noindent where the thermodynamic variables
$\beta,{\bf v}_n$ and $\mu$, together with $U$, all depend on
${\bf r}$ and $t$. When (\ref{eq6}) is substituted into
(\ref{eq5}), the collision
integral vanishes by virtue of the local equilibrium form of the
%distribution function, and leads to 
distribution function. Taking moments of the resulting equation
with respect to {\it 1}, $p_\mu$ and $p^2$, we obtain the
closed set of equations\cite{newref10}
%an equation which can be solved by the method of moments.
%Specifically, taking moments with respect to {\it 1}, $p_\mu$ 
%and $p^2$, we obtain the following set of 
%equations\cite{newref10}
\begin{eqnarray}
{\partial \tilde n\over\partial t}
+\bbox{\nabla}\cdot(\tilde n {\bf v}_n) &=& 0 \nonumber\\
m \tilde n \left [ {\partial {\bf v}_n\over \partial t} 
+ ({\bf v}_n\cdot\nabla){\bf v}_n \right ] &=& -\
\bbox{\nabla} \tilde P 
-\ \tilde n\bbox{\nabla} U \nonumber \\
{\partial \tilde\epsilon \over \partial t} + {5\over
3}\bbox{\nabla}\cdot (\tilde\epsilon{\bf v}_n)  &=&
\ {\bf v}_n\cdot\bbox{\nabla} \tilde P \,.
\label{eq7}
\end{eqnarray}
The non-condensate density $\tilde n({\bf r},t)$ is given by
\begin{equation}
\tilde n({\bf r},t) \equiv\int {d^3 p\over h^3}
f_0({\bf r},{\bf p},t) = {1\over \Lambda^3}
g_{3/2}(z({\bf r},t))\ ,
\label{eq8}
\end{equation}
\noindent with $z({\bf r},t) \equiv e^{\beta({\bf r},t)[\mu({\bf
r},t) - U({\bf r},t)]}$ and $\Lambda({\bf r},t) = (2\pi\hbar^2/$
$mk_B T({\bf r}, t))^{1/2}$. The quantity $\tilde \epsilon$ is the
non-convective part of the kinetic energy density defined with
${\bf v}_n = 0$ in (\ref{eq6}).  Similarly, $\tilde P = {2\over 3} 
\tilde \epsilon$ is the kinetic contribution to
the local equilibrium pressure defined by 
\begin{eqnarray}
\tilde P({\bf r},t) &\equiv& \int {d^3 p\over h^3}
{p^2\over 3m} f_0({\bf r},{\bf p},t)\big \vert_{{\bf v}_n = 0}
\nonumber \\
&=& {1\over\beta\Lambda^3} g_{5/2} (z({\bf r},t)) \,.
\label{eq9}
\end{eqnarray}
Eqs. (\ref{eq3}) and (\ref{eq7}) constitute our full set of
nonlinear
hydrodynamic equations for a trapped Bose gas at finite
temperatures. 

At the level of approximation we are considering, the condensate and
non-condensate satisfy separate continuity equations. 
Combining these two equations gives the
expected two-fluid continuity equation

\begin{equation}{\partial n\over\partial t} = -\bbox{\nabla}\cdot
{\bf j}\ ,
\label{eq10}
\end{equation}

\noindent where $ n\equiv\tilde n + n_c$ and $
{\bf j} \equiv\tilde n {\bf v}_n + n_{c}{\bf v}_c$.
We identify the normal fluid density with $\tilde n({\bf r},t)$ and the
superfluid density with $n_c({\bf r},t)$. The former identification is
supported by noting that (\ref{eq8}) can be expressed
equivalently as
\begin{equation}
\tilde n({\bf r},t) = -\int {d^3 p\over h^3} {p^2\over 3m}
{\partial
f_0(\varepsilon_p)\over \partial \varepsilon_p}\ ,
\label{eq15}
\end{equation}
where $\varepsilon_p({\bf r},t) \equiv{p^2\over 2m} + 
U({\bf r},t) -\mu({\bf r},t)$ is the excitation energy. This is
the usual Landau formula for the normal fluid
density\cite{landau41,noz90}.

The linearized version of (\ref{eq3}) and (\ref{eq7}) allows 
one to consider small amplitude oscillations about equilibrium. 
The equilibrium condensate
wavefunction is determined by the solution of
\begin{eqnarray}
\hat{\cal H}_0({\bf r})\Phi_0({\bf r}) 
\equiv\Big [&-&{\hbar^2 \nabla^2\over 
2m} +U_{ext}({\bf r}) + 2g\tilde n_0({\bf r}) \nonumber \\
&+&gn_{c0}({\bf r})\Big ] \Phi_0({\bf r}) =\mu_0\Phi_0({\bf r})\  ,
\label{eq11}
\end{eqnarray}
with $n_{c0}({\bf r}) = |\Phi_0({\bf r})|^2$. The equilibrium 
non-condensate density $\tilde n_0({\bf r})$ is given by (\ref{eq8})
with the 
equilibrium fugacity defined as $z_0 = e^{\beta_0[\mu_0 - U_0({\bf
r})]}$, with $U_0({\bf r}) = U_{ext}({\bf r}) + 2gn_0({\bf r})$.
Eq.(\ref{eq11}) and the equilibrium version of
(\ref{eq8}) must be solved self-consistently.

The linearization of (\ref{eq3}) around equilibrium leads to the 
condensate equations
\begin{eqnarray} 
{\partial \delta n_c\over \partial t}&=& -\bbox{\nabla}
\cdot (n_{c0}\delta {\bf v}_c) \nonumber \\
m {\partial \delta {\bf v}_c\over \partial t} &=&
-\bbox{\nabla}\delta \phi\ ,
\label{eq12}
\end{eqnarray}
where
\begin{eqnarray}
\delta \phi({\bf r},t) \equiv {1\over \vert\Phi_0({\bf r})\vert}
[\hat {\cal H}_0({\bf r}) &-& \mu_0] \delta \vert\Phi({\bf r},t)\vert 
\nonumber \\ &+& g\delta n_c({\bf r},t) 
+ 2g\delta \tilde n({\bf r},t)\ .
\label{eq13}
\end{eqnarray}
In arriving at this result, we have noted that $\nabla \phi_0({\bf
r}) =
0$ in equilibrium.
Similarly, the linearization of (\ref{eq7})
leads 
to the equations
\begin{eqnarray}  
{\partial\delta\tilde n\over\partial t} &=& -\bbox{\nabla} \cdot
(\tilde n_0 \delta {\bf v}_n)\nonumber\\
m\tilde n_0{\partial\delta{\bf v}_n\over \partial t} &=&
-\bbox{\nabla}
\delta\tilde P -\delta\tilde n\bbox{\nabla} U_0 - 2g\tilde
n_0\bbox{\nabla} (\delta\tilde n +\delta n_c) \nonumber \\
{\partial\delta\tilde P\over \partial t} &=& - {5\over
3}\bbox{\nabla}\cdot
(\tilde P_0\delta {\bf v}_n) + {2\over 3} \delta {\bf
v}_n \cdot\bbox{\nabla} \tilde P_0\,,
\label{eq14}
\end{eqnarray}
where $\tilde P_0({\bf r})$ is the equilibrium kinetic pressure 
which satisfies $\bbox{\nabla}\tilde P_0 = -\tilde 
n_0\bbox{\nabla} U_0$.
Above $T_{BEC}$, these equations reduce to those of Ref.
\protect\cite{gri97} if we ignore the effect of interactions
$(g=0)$.  

For its intrinsic interest, and in
order to better understand the implications of our two-fluid
equations, we now consider 
the limit of a {\it homogeneous} system $(U_{ext}({\bf r})=0)$. 
In this special case, (\ref{eq11}) yields a uniform condensate with
the
chemical potential having the Thomas-Fermi (TF) value
$\mu_0 = 2g\tilde n_0 + gn_{c0}$. Taking all equilibrium
quantities
to be spatially-independent, and noting that the first term on the
right 
hand side of (\ref{eq13}) can be neglected
in the long-wavelength limit, the two velocity equations
reduce to
\begin{eqnarray}
m {\partial \delta{\bf v}_c\over\partial t} &=&
-2g\bbox{\nabla}\delta\tilde n -
g\bbox{\nabla}\delta n_c\nonumber\\
m\tilde n_0{\partial\delta{\bf v}_n\over \partial t} &=&
-\bbox{\nabla}\delta\tilde P -
2g\tilde n_0\bbox{\nabla}\delta\tilde n-2g\tilde
n_0\bbox{\nabla}\delta n_c\ ,
\label{eq16}
\end{eqnarray}

\noindent  with

\begin{equation}
{\partial\delta\tilde P\over\partial t} = -{5\over
3}\tilde P_0\bbox{\nabla}\cdot\delta{\bf v}_n \ .
\label{eq17}
\end{equation}
The two equations in (\ref{eq16}) can be combined to give
\begin{eqnarray}
{\partial\over \partial t}\delta {\bf j}& =
&-\bbox{\nabla}\delta\tilde P -
2g(n_{c0} + \tilde n_0)\bbox{\nabla} \delta\tilde n-g(n_{c0} +
2\tilde n_0)\bbox{\nabla}\delta n_c \nonumber \\
&\equiv& -\bbox{\nabla}\delta P\ . \label{eq18}
\end{eqnarray}

\noindent One can verify that (\ref{eq18}) is consistent with the
following expression for the {\it total} local thermodynamic
pressure 

\begin{equation}
P=\tilde P + {1\over 2} g[n^2+2n\tilde n-\tilde n^2]\ ,
\label{eq19}
\end{equation}
which is the equation of state at the level of approximation we are
considering. Within the same approximation,
the total internal energy density is given by

\begin{eqnarray}
\epsilon({\bf r},t) &=&\tilde \epsilon({\bf r},t) + {1\over
2}g\langle
\hat\psi^\dagger ({\bf r})\hat\psi^\dagger ({\bf r})\hat\psi({\bf
r})\hat\psi({\bf r})\rangle_t\nonumber\\
&\simeq&\tilde\epsilon({\bf r},t) +{1\over 2} g[n^2+2n\tilde n-\tilde
n^2]\ .
\label{eq20}
\end{eqnarray}

These results and the equilibrium thermodynamic relation
\begin{equation}
\epsilon + P = sT+\mu n\ ,
\label{eq22}
\end{equation}
allow us to identify the entropy density $s$.
Using $\mu_0 = g(2\tilde n_0 + n_{c0})$
together with (\ref{eq19}) and (\ref{eq20}), (\ref{eq22})
gives
\begin{equation}
s_0T_0 ={5\over 2} \tilde P_0 + g n_{c0}\tilde n_0\ .
\label{eq23}
\end{equation}
Here $\tilde P_0$ is the equilibrium kinetic pressure
defined in (\ref{eq9}), 
with $z \equiv z_0 = e^{\beta_0(\mu_0- 2gn_0)} = e^{- \beta
gn_{c0}}$.  Using the local equilibrium expression for  $\tilde P$
in (\ref{eq9}), one finds that the fluctuation in the total
pressure $P$ is given by
\begin{equation}
\delta P = s_0\delta T +\tilde n_0\delta\mu + g n_{c0} 
(2\delta\tilde n +\delta n_c)\ ,
\label{eq24}
\end{equation}
with $s_0$ defined by (\ref{eq23}).  Comparing
this to the thermodynamic relation $\delta P = s_0 \delta T +
n_0\delta\mu$, we arrive at

%\delta\epsilon &=& T_0\delta s +\mu_0\delta n \label{eq33}\\

\begin{equation}
\delta\mu = \delta (2g \tilde n + g n_c)\ .
\label{eq25}
\end{equation}
This result confirms (in the case of a uniform equilibrium
density) that $\delta \phi({\bf r},t)$ in (\ref{eq12}) and
(\ref{eq13}) is indeed the fluctuation in the local
chemical potential.  More generally, we have verified that
(\ref{eq3}) is equivalent to the key Landau equation for superfluid
flow \protect\cite{landau41,noz90,hoh65}

\begin{equation}
m\left [{\partial{\bf v}_S\over \partial t} +{1\over 2}
\bbox{\nabla}{\bf v}_S^2\right ] = - \bbox{\nabla}\mu ({\bf r},t)\
.
\label{eq26}
\end{equation}

Finally, using (\ref{eq20}) we find

\begin{equation}
\delta\epsilon = {3\over 2} \delta\tilde P + gn_{c0}
(2\delta\tilde n +\delta n_c) + 2g\tilde n_0\delta n\ .
\label{eq27}
\end{equation}

\noindent Inserting this result into the thermodynamic relation 
$\delta\epsilon = T_0\delta s +\mu_0\delta n$, a simple calculation
gives

\begin{equation}
T_0\delta s = {3\over 2}\delta\tilde P +
gn_{c0}\delta \tilde n\ .
\label{eq28}
\end{equation}

\noindent Taking the time-derivative,  and using (\ref{eq17})
and the continuity equation for $\delta \tilde n({\bf r},t)$, we
obtain  the linearized form of Landau's entropy
conservation equation\protect\cite{landau41,noz90,hoh65}

\begin{equation}
{\partial\delta s\over \partial t} = -
s_0\bbox{\nabla}\cdot\delta{\bf v}_n\
.
\label{eq29}
\end{equation}

In the uniform case, Eqs. (\ref{eq16}) and (\ref{eq17}) are easily
solved to give the expected first and
second sound phonon modes. In contrast to superfluid $^4$He, second
sound in a gas involves a condensate oscillation largely uncoupled
from the
non-condensate, with a velocity given by $u_2=(gn_{c0}/m)^{1/2}$.

As a specific application to a trapped Bose gas,
we consider the center-of-mass mode solution
for  an anisotropic parabolic potential
$U_{ext}({\bf r}) = {1\over 2}m(\omega_x^2 x^2 + \omega_y^2 y^2 +
\omega_z^2 z^2)$. On the basis of the generalized Kohn theorem 
\protect\cite{newref11}, one expects three modes in which the 
gas oscillates rigidly along each of the principal directions at the
appropriate trap frequency $\omega_i$. 
Denoting the displacement of the gas by $\bbox{\eta}(t)$,
the condensate and non-condensate densities {\it both} behave as 
$n_0({\bf r} - \bbox{\eta}(t))$,
giving a density fluctuation $\delta n({\bf r},t) = - 
\nabla n_0({\bf r}) \cdot \bbox{\eta}(t)$, and a velocity field 
${\bf v}({\bf r},t) = \dot{ \bbox{\eta}}(t)$ which
is spatially independent. One can easily verify that our linearized
equations (\ref{eq12}) and (\ref{eq14}) admit a solution of this
kind
with both the condensate and non-condensate having identical 
displacements
$\bbox{\eta}(t) =\bbox{\eta}_0\cos \omega_i t$. Thus, in
contrast to the static HFP used in Ref.\protect\cite{hut97}, the
dynamic
HFP theory given by (\ref{eq1}) is consistent with the generalized
Kohn theorem. 

An estimate of the frequency of the out-of-phase 
dipole mode in a trapped gas can be based on the
intuitive idea that the condensate and non-condensate oscillate
rigidly
against each other. The problem is then equivalent to a
two-particle
problem with masses $M_c = mN_c$ and $M_n = m\tilde N$, each
confined in the parabolic potential and coupled together by
a spring with force
constant $k$. Since the interaction energy between the
two components is $E_{int} = 2g \int d{\bf r}\, n_c({\bf r}) \tilde
n({\bf r})$, $k$ can be determined by considering
small displacements of the two components along the $i$-th
direction. The equations of motion for the two coupled masses can
then 
be solved, giving the in-phase center-of-mass mode at the
frequency $\omega_i$ discussed above,
and an out-of-phase mode at the frequency
\begin{equation}
\Omega_i^2 = \omega_i^2 - 2g {M_c+M_n\over M_cM_n} \int d{\bf r}\,
{\partial n_{c0}({\bf r}) \over \partial x_i}\, {\partial \tilde 
n_0({\bf r}) \over \partial x_i}\,.
\label{eq31}
\end{equation}
This result is confirmed by a more detailed 
treatment\protect\cite{zaremba} in which the hydrodynamic 
equations (\ref{eq12}--\ref{eq14}) are cast into the form of a
variational principle. In Fig. 1 we show the in- and out-of-phase
mode frequencies as a function of temperature for a gas trapped
in an isotropic parabolic potential.  We also show the relative
amplitudes of the condensate and non-condensate oscillations
which satisfy the condition $M_c\eta_c + M_n\eta_n =0$, corresponding
to the center of mass being stationary for any temperature.
A measurement of these amplitudes would therefore determine 
directly the ratio $N_c/\tilde N$ of the two fluid components.
This mode of the trapped Bose gas is the analogue of the usual 
second sound mode\protect\cite{noz90} in bulk superfluid $^4$He 
(for which $\rho_S{\bf v}_S +\rho_N {\bf v}_N =0$).

In summary, we have given a microscopic derivation of the
hydrodynamic two-fluid equations for a trapped weakly-interacting
Bose gas, as summarized by Eqs. (\ref{eq12}) -- (\ref{eq14}). Our 
analysis shows how such a  two-fluid description arises naturally
from the existence of a macroscopic condensate wavefunction
which is coupled to the non-condensate
atoms. We have also shown explicitly how our linearized two-fluid
equations
reduce to the well-known equations (\ref{eq3}), (\ref{eq10}),
(\ref{eq18}) and (\ref{eq29}) of
Landau\protect\cite{landau41,noz90} when applied to
a uniform Bose-condensed gas. Finally, we have shown that
our equations are consistent with the generalized Kohn theorem
and obtained the
out-of-phase mode analogue of second-sound in a trapped Bose gas.
A more complete discussion of the two-fluid dynamics will be
given elsewhere\protect\cite{zaremba}.

%\acknowledgements
A.G. would like to thank Wen-Chin Wu for discussion at the early
stages of this work and E.Z. would like to thank Brian W. King
for help with the numerical calculations. 
This research was supported by grants from
NSERC of Canada.

%%%%%%

\begin{figure}
\caption{
(a) Mode frequencies for the in-phase (solid dots) and
out-of-phase (open dots) dipole modes vs. temperature
for 2000 Rb atoms in an isotropic parabolic trap (see
Ref. [5] for values of the physical parameters used).
(b) Condensate ($\eta_c$) and non-condensate ($\eta_n$) 
amplitudes for the out-of-phase dipole mode.
(c) Fraction of atoms in the condensate as a function of
temperature.
}
\label{figure1}
\end{figure}


\begin{references}


\bibitem{tisza38}L. Tisza, Nature (London), {\bf 141}, 913 (1938);
J. Phys. Radium, {\bf 1}, 165 (1940).

\bibitem{landau41}L.D. Landau, J. Phys. U.S.S.R. {\bf 5}, 71
(1941).

\bibitem{noz90}P. Nozi\`eres and D. Pines, {\it The Theory of
Quantum
Liquids}, Vol. 2 (Addison-Wesley, Redwood City, CA, 1990), Chs. 7
and 10.

\bibitem{hoh65}P.C. Hohenberg and P.C. Martin, Ann. Phys. (N.Y.),
{\bf 34}, 291 (1965).

\bibitem{hut97}D.A.W. Hutchinson, E. Zaremba and A. Griffin, \prl
{\bf 78}, 1842 (1997).

\bibitem{griffin96}A. Griffin, \prb {\bf 53}, 9341 (1996).

\bibitem{gri97}A. Griffin, W.C. Wu and S. Stringari, \prl {\bf 78},
1838 (1997).

\bibitem{gio}S. Giorgini, L. Pitaevskii and S. Stringari, Journ.
Low Temp. Phys., to be published; cond-mat/9704014. These
authors show that the particle-like excitation energy we use is
valid to quite low temperatures in a trapped gas.

\bibitem{KB} L. P. Kadanoff and G. Baym, {\it Quantum
Statistical Mechanics} (W.A. Benjamin, N.Y., 1962), Ch. 6.

\bibitem{ref11}T.R. Kirkpatrick and J.R. Dorfman, Journ. Low Temp.
Phys. {\bf 58}, 304 (1985); {\it ibid.} 399 (1985).
While quite different in detail, our
work may be viewed as a generalization of these papers to the case of
a trapped Bose gas.

\bibitem{newref10}
Eq.(58) in the first paper of Ref.\protect\cite{ref11} 
is a kinetic equation for
excitations in the local rest frame of the condensate. We have
verified that this equation leads to the same set of hydrodynamic
equations as in (\ref{eq7}) if excitations in and out of the
condensate are ignored.

\bibitem{newref11}J.F. Dobson, \prl {\bf 73}, 7244 (1994).

\bibitem{zaremba}E. Zaremba and A. Griffin, to be published.


%\bibitem{stringari96}For a similar calculation at $T=0$, see S.
%Stringari, \prl {\bf 77}, 2360 (1996).

%\bibitem{newref12}T.D. Lee and C.N. Yang, Phys. Rev. {\bf 112},
%559
%(1958).

%\bibitem{ref12}T.D. Lee and C.N. Yang, Phys. Rev. {\bf 113}, 1406
%(1959) .



\end{references}
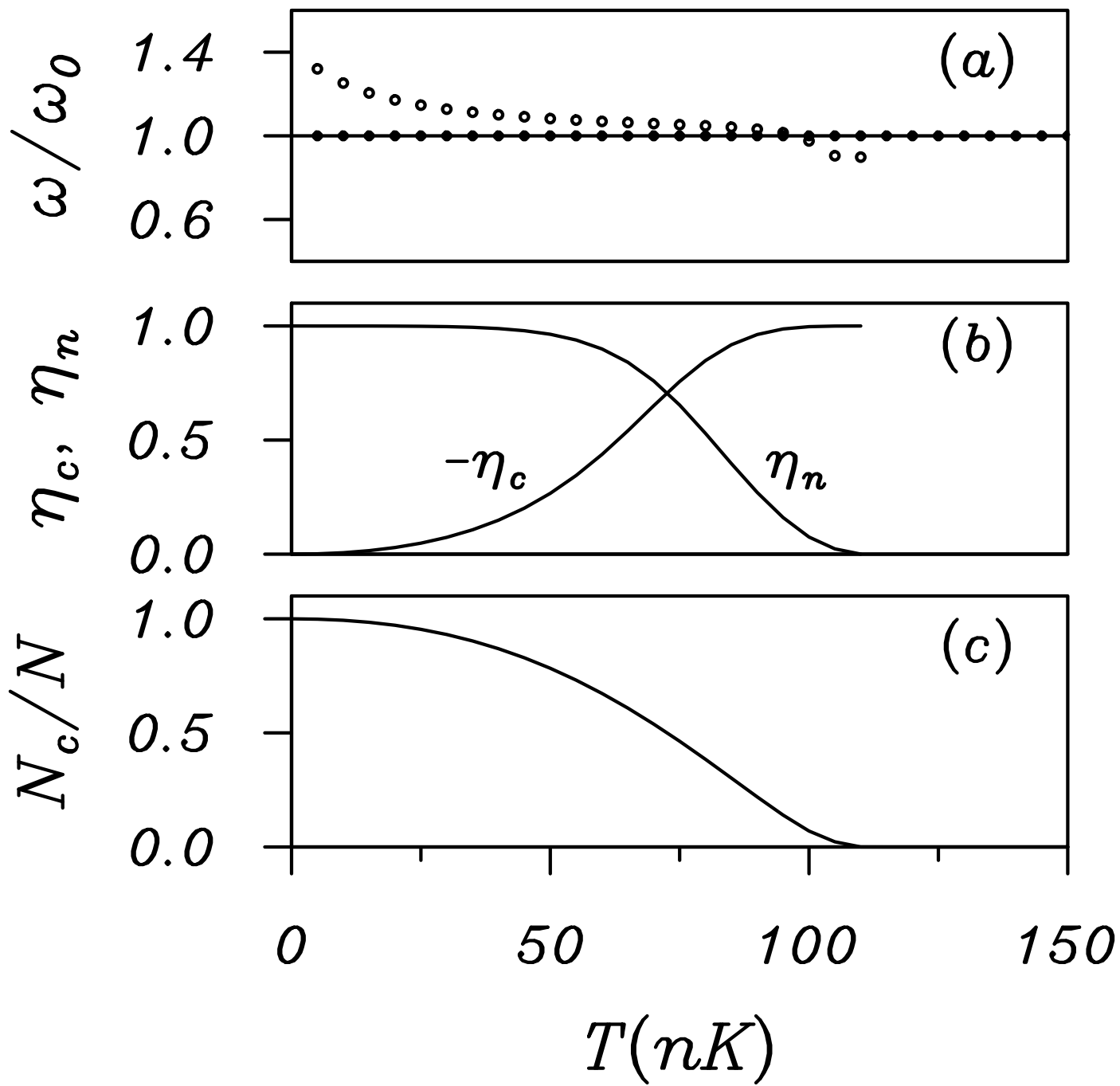
\end{document}